\documentclass[11pt]{article}
\usepackage{graphics}
\usepackage{lscape,epsfig}
\usepackage{amsmath}
\usepackage{amsfonts}
\usepackage{amssymb}
\usepackage{natbib}
\usepackage{enumerate}
\usepackage[title,titletoc,toc]{appendix}
\usepackage{color}
\usepackage{tikz}
\usepackage[normalem]{ulem}
\usepackage{mathtools}
\usepackage{authblk}
\usepackage{hyperref}

\usetikzlibrary{arrows,positioning,shapes.geometric}
\definecolor{red}   {RGB}{180,0,0}
\definecolor{gray10}{rgb}{0.1,0.1,0.1}
\definecolor{gray20}{rgb}{0.2,0.2,0.2}
\definecolor{gray30}{rgb}{0.3,0.3,0.3}
\definecolor{gray40}{rgb}{0.4,0.4,0.4}
\definecolor{gray50}{rgb}{0.5,0.5,0.5}
\definecolor{gray60}{rgb}{0.6,0.6,0.6}
\definecolor{gray80}{rgb}{0.8,0.8,0.8}
\definecolor{gray90}{rgb}{0.9,0.9,.9}
\definecolor{gray95}{rgb}{0.95,0.95,.95}
\definecolor{gray96}{rgb}{0.96,0.96,.96}
\definecolor{sgGreen} {RGB}{20, 180, 50}
\newcommand{\dashedrightarrow}[1][2pt]{  \settowidth{\@tempdima}{$\rightarrow$}\rightarrow  \makebox[-\@tempdima]{\hskip-1.5ex\color{white}\rule[0.5ex]{#1}{1pt}}  \phantom{\rightarrow}}
\newtheorem{innercustomgeneric}{\customgenericname}
\providecommand{\customgenericname}{}
\newcommand{\newcustomtheorem}[2]{%
  \newenvironment{#1}[1]
  {%
   \renewcommand\customgenericname{#2}%
   \renewcommand\theinnercustomgeneric{##1}%
   \innercustomgeneric
  }
  {\endinnercustomgeneric}
}

\newcustomtheorem{customthm}{Theorem}
\newcustomtheorem{customlemma}{Lemma}
\newcustomtheorem{customcorollary}{Corollary}

\DeclarePairedDelimiterX\Basics[1](){ #1}

\def\appendix{\par \setcounter{section}{0} \def\thesection{A}}
\begin{document}

\title{More testing or more disease? \\
A counterfactual approach to explaining observed increases in positive tests over time }
\author[1]{Jessica G. Young\thanks{Corresponding Author: jyoung@hsph.harvard.edu}}

\author[1]{Sarah J. Willis}
\author[2]{Katherine Hsu}
\author[1,3]{Michael Klompas}
\author[1]{Julia L. Marcus}


\affil[1]{Department of Population Medicine, Harvard Medical School and Harvard Pilgrim Health Care Institute, MA, USA}
\affil[2]{Massachusetts Department of Public Health, MA, USA}
\affil[3]{Department of Medicine, Brigham and Women's Hospital, MA, USA}



\date{}
\setcounter{Maxaffil}{0}
\renewcommand\Affilfont{\itshape\small}

\maketitle

\begin{abstract}
Observed gonorrhea case rates (number of positive tests per 100,000 individuals) increased by 75\% in the United States between 2009 and 2017, predominantly among men.  However, testing recommendations by the Centers for Disease Control and Prevention (CDC) have also changed over this period with more frequent screening for sexually transmitted infections (STIs) recommended among men who have sex with men (MSM) who are sexually active.  In this and similar disease surveillance settings, a common question is whether observed increases in the overall proportion of positive tests over time is due only to increased testing of diseased individuals, increased underlying disease or both.   By placing this problem within a counterfactual framework, we can carefully consider untestable assumptions under which this question may be answered and, in turn, a principled approach to statistical analysis.  This report outlines this thought process.
\end{abstract}

\section{Introduction}\label{intro}
Observed gonorrhea case rates (number of positive tests per 100,000 individuals) increased by 75\% in the United States between 2009 and 2017 \citep{cdc2018}, predominantly among men.  However, testing recommendations by the Centers for Disease Control and Prevention (CDC) have increasingly emphasized more frequent screening for sexually transmitted infections (STIs) among men who have sex with men (MSM) who are sexually active \citep{cdc2010,cdc2015}. In 2014, CDC also recommended STI screening every six months for users of HIV preexposure prophylaxis (PrEP)\citep{cdcprep} and updated their guidelines to recommend screening PrEP users every three months in 2017\citep{cdcprep2017}.  

In this and similar disease surveillance settings, a common question is whether observed increases in the overall proportion of positive tests over time is due to increased targeted screening efforts (i.e. increased chance of identifying disease in sick individuals), increased underlying disease (the population is getting sicker over time) or both.  Answering this question requires knowledge of quantities that cannot be generally identified (and, in turn, estimated) without untestable assumptions.  We can rely on results from the causal inference literature \citep{hernanfail,pearldag,causalbook} and the related missing data literature \citep{littlerubin}, to explicitly consider these untestable assumptions and, in turn, inform an appropriate analytic approach.  

\section{The observed data and observable quantities}\label{obsdata}
Consider a time period of interest divided into equally spaced intervals.  Define $n_t$ as the total number of individuals in the study population of interest (e.g. men in the United States) at the start of interval $t$.  We allow that each interval $t$ study population may consist of different individuals with some individuals members of the population at multiple times.  

Assume we observe the following for each individual $i=1,\ldots,n_t$ in each interval $t$ study population.  Suppressing the individual $i$ subscript, denote $S_t$ as an indicator of whether the individual is tested for a disease (e.g. gonorrhea) in that interval and $Z_t$ as a vector of pre-test disease risk factors for that individual (e.g. number of previous STIs and STI tests, previous HIV tests, age, race, PrEP use).  Denote $Y_t$ as an indicator of whether an individual has that disease in interval $t$. For individuals without a test in interval $t$ ($S_t=0$), disease status $Y_t$ is unobserved.  We rely on the temporal order assumption ($Z_t,S_t,Y_t)$.

We make the simplifying assumptions that if $S_t=1$ then $Y_t$ is known, no subject in each interval $t$ study population dies or is lost to follow-up before the end of the interval and screening/testing does not affect disease status within an interval (e.g. it cannot affect behavior change).  These assumptions could be guaranteed for increasingly short intervals (e.g. hours or days) but are not guaranteed for longer intervals (e.g. years).  Finally, we assume no measurement error, including 100\% sensitivity and specificity of the test.  The ideas presented here can in principle be extended to accommodate more complicated settings.  For example, see \cite{xaviscreen,obsplans,sonjascreen,romainnde}.

In the current setting, we require no untestable assumptions in order to confirm whether the population chance of receiving a test in any interval $t$, which we can denote $\Pr[S_t=1]$, is increasing with time $t$.  In each interval $t$ this chance can be estimated via the sample proportions $\frac{n_{s,t}}{n_t}$ where $n_{s,t}=\sum_{i=1}^{n_t}S_{i,t}$ (the number of individuals in the population who received a test).  These proportions might be plotted as a function of interval $t$ to observe trends over time which in some cases may reasonably be summarized by a regression model.  A simple model is 
$\Pr[S_t=1]=\mbox{exp}\{\beta_0+\beta_1t\}$.  Provided this regression model is correctly specified, a value of $\beta_1>0$ would be consistent with an increasing trend.  This model is simple because it assumes $\frac{\Pr[S_t=1]}{\Pr[S_{t-1}=1]}$ is the same for all $t$.  More complex models that allow more flexible functions of $t$ can also be considered (e.g. higher order polynomials, splines).  A completely flexible model would include separate indicator terms for each interval in the follow-up.  This would allow $\frac{\Pr[S_t=1]}{\Pr[S_{t-1}=1]}$  to differ for all $t$.
 Generalized estimating equations \citep{gee} might be used to estimate model coefficients and construct 95\% confidence intervals that account for the fact that some individuals might contribute to more than one interval $t$ population.  Analogously, we can directly observe trends within subsets of the population defined by a particular level of $Z_t$; i.e., trends in $\Pr[S_t=1|Z_t=z_t]$ for some level $Z_t=z_t$.  This might be achieved by stratifying regression models on different levels of $Z_t$.  More generally, it will require additional regression model assumptions.   

We similarly require no untestable assumptions in order to infer whether the prevalence of disease \textsl{conditional on being tested}, which we can denote $\Pr[Y_t=1|S_t=1]$, is increasing with time $t$.  In each interval $t$, this chance can also be estimated via sample proportions, here computed as $\frac{n_{y,s,t}}{n_{s,t}}$ where $n_{y,s,t}=\sum_{i=1}^{n_t}Y_{i,t}S_{i,t}$ (the number of individuals in the population who had a positive test).   Similarly, regression models might be used for summarizing the trends and robust methods used for variance estimation, possibly with further conditioning on $Z_t$.  

The overall chance of a positive test is the joint population chance that $Y_t=1$ and $S_t=1$ (having disease and having a test) or $\Pr[Y_t=1,S_t=1]$.  This quantity is also observed in this setting and could be estimated via the sample proportion $\frac{n_{y,s,t}}{n_{t}}$, the number of individuals in interval $t$ with disease and a test divided by the total in the population at $t$.  We would like to understand whether increases in this quantity are explained by increased testing in diseased individuals or increased underlying disease.

\section{More disease? }\label{estimand}
An answer to our question minimally requires knowledge of a quantity that we do not directly observe in any interval $t$: the prevalence of disease in the whole interval $t$ population for each $t$.  This quantity is not directly observable because disease status $Y_t$ is not observed if an individual does not receive a test ($S_t=0$).  Relative to the observed data, we can understand this question in terms of time trends in a counterfactual probability.

Specifically, let $Y^{s_t=1}_t$ denote the disease indicator we would have observed for an individual in the interval $t$ population had, possibly contrary to fact, that individual received a test in interval $t$.  In turn, the underlying population disease prevalence for interval $t$ had we implemented a hypothetical intervention to ensure everyone in the population was tested is
\begin{equation}
\Pr[Y^{s_t=1}_t=1].\label{truepop}
\end{equation} 
Unlike the quantities considered in the previous section, we cannot directly assess trends in $\Pr[Y^{s_t=1}_t=1]$ as a function of time $t$ as these probabilities cannot be known without assumptions.  We allow that interest might also be in trends in subsets of the population defined by a subset $V_t$ of the covariates $Z_t$ 
\begin{equation}
\Pr[Y^{s_t=1}_t=1|V_t]\label{truesub}
\end{equation}
where a special case is $V_t=Z_t$ (i.e. the subset is the whole set of covariates).  In the next section we consider untestable assumptions under which these generally unobservable quantities and their corresponding trends in $t$ can be identified using only the observed data. 

\subsection{Identifying assumptions}\label{assumptions}
Suppose the following assumptions hold for each interval $t$ population:
\begin{enumerate}
\item Conditional exchangeability: $Y^{s_t=1}\coprod S_t|Z_t$\\
This assumption requires that, within levels of measured baseline covariates, whether or not someone receives a test in interval $t$ is independent of his underlying disease status that we would observe had he been screened.  Conditional exchangeability may be alternatively referenced as the assumption of \textsl{no unmeasured selection bias} with $Z_t$ referred to as the \textsl{measured selection factors} \citep{causalbook} and also coincides with a missing at random (MAR) assumption \citep{littlerubin}.  Conditional exchangeability is guaranteed under an intervention on $S_t$ where the investigators flipped a weighted coin dependent only on $Z_t$ to determine who does and does not receive a test.  However, in the absence of intervention on $S_t$, this assumption is not guaranteed to hold.  
\item Positivity: If $\Pr[Z_t=z_t]>0$ then $\Pr[S_t=1|Z_t=z_t]>0$\\
This assumption requires that, if a particular level of the baseline covariates $Z_t=z_t$ is possible to observe in the interval $t$ population, then it must also be possible for individuals with that level of the baseline covariates to receive a test.
\item Consistency: If $S_t=1$ then $Y^{s_t=1}=Y_t$\\
This assumption states that, if a subject is observed to receive a test, then his observed disease status is equal to his counterfactual disease status. Consistency requires well-defined interventions \citep{waterkill}, implying the so-called stable unit treatment value assumption (SUTVA)\citep{rubinsutva} which includes the assumptions of ``no multiple versions of treatment'' and ``no interference between units'' (i.e., one individual's treatment does not affect a different individual's counterfactual outcome).  Consistency holds under the simplifying assumption that screening cannot affect disease status within interval $t$ but is not generally guaranteed. 
\end{enumerate}

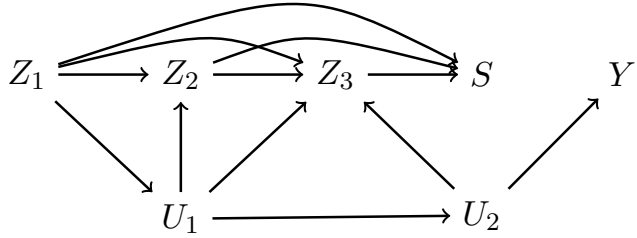
\begin{figure}[tbp]
\begin{center}
\scalebox{1.2}{
\begin{tikzpicture}

         \draw[thick,->] (0,-5) node[left] (Z2) {$Z_2$} -- (1,-5) node[right] (Z3) {$Z_{3}$};         
        \node[right=1cm of Z3] (S) {$S$};
        \node[left=1cm of Z2] (Z1) {$Z_{1}$};
         \node[right=1cm of S] (Y) {$Y$};
         \node[below=1cm of S] (U2) {$U_2$};
          \node[below=1cm of Z2] (U1) {$U_1$};
          \draw[thick,->] (U2) -- (Z3);  
          \draw[thick,->] (U2) -- (Y);   
           \draw[thick,->] (Z3) -- (S);   
            \draw[thick,->] (Z1) -- (Z2);
             \draw[thick,->] (U1) -- (Z2);
              \draw[thick,->] (U1) -- (Z3);
               \draw[thick,->] (U1) -- (U2);
                \draw[thick,->] (Z1) -- (U1);
               \draw[thick,->] (Z1) .. controls (0,-4.5) .. (Z3);  
                 \draw[thick,->] (Z1) .. controls (1,-4.0) .. (S);  
                  \draw[thick,->] (Z2) .. controls (1,-4.5) .. (S);  
                \end{tikzpicture}}
\end{center}
\caption{A causal DAG representing observed data generating assumptions under which we may identify the counterfactual disease prevalence had all individuals been tested $\Pr[Y^{s_t=1}_t=1]$. No unblocked backdoor paths (noncausal paths) connecting testing $S$ to disease $Y$ conditional only on measured covariates $Z_1,Z_2,Z_3$}
\label{dag}
\end{figure}

Figure \ref{dag} depicts a causal directed acyclic graph (DAG) representing a possible data generating assumption on the observed data in interval $t$ described in Section \ref{obsdata} with the $t$ subscripts suppressed.  In this graph, assume $Z_1$ (e.g. race, age), $Z_2$ (e.g. a prior rectal STI test) and $Z_3$ (STI symptoms) comprise the vector of measured covariates $Z$ while $U_1$ (e.g. previous condomless sex) and $U_2$ (e.g. previously undiagnosed disease) represent unmeasured risk factors for current disease status in interval $t$.  Paths on the graph connecting two nodes via only directed arrows represent causal paths.  For example, the path $Z_3\rightarrow S$ represents the assumption that disease symptoms could affect an individual's chance of receiving a test for the disease.  By contrast, the path $Z_3\leftarrow U_2\rightarrow Y$  is an example of an ``unblocked backdoor'' (noncausal) path, which represents the assumption that disease symptoms are associated with current disease status via their common cause, ``previously undiagnosed disease''.  The assumption of conditional exchangeability, or no unmeasured selection bias, is consistent with the assumptions of Figure \ref{dag} by the absence of any unblocked backdoor paths connecting $S$ and $Y$ conditional on $Z_1$, $Z_2$ and $Z_3$ only \citep{pearldag,swigs}.   This assumption would fail if symptoms $Z_3$ were unmeasured (and therefore not a component of the measured covariate vector $Z$).  Alternatively, this assumption would fail if previous condomless sex $U_1$ directly affected the individual's decision to receive a test (which would be represented by adding an arrow from $U_1$ into $S$).

Given conditional exchangeability, consistency and positivity, it is straightforward to show that our counterfactual probability of interest $\Pr[Y^{s_t=1}_t=1]$ can be written as a function of only the observed data:
\begin{equation}
\sum_{z_t}\Pr[Y_t=1|S_t=1,Z_t=z_t]\Pr[Z_t=z_t]\label{gform}
\end{equation}
The expression (\ref{gform}) is a special case of Robins's g-formula \citeyear{robinsfail}.  The expression can be most simply understood as a weighted sum of the chance of having disease \textsl{conditional on receiving a test and covariate level $Z_t=z_t$} over all possible levels, weighted by the chance of that covariate level occurring in the interval $t$ population.  When $Z_t$ contains any continuous components, the sum would be replaced with an integral.  

Note that, in the special case where \textsl{marginal exchangeability} holds, rather than \textsl{conditional exchangeability}, such that  $Y^{s_t=1}\coprod S_t|Z_t$ for $Z_t=\varnothing$ (the empty set), then the expression (\ref{gform}) reduces simply to the chance of a positive test amongst those tested  $\Pr[Y_t=1|S_t=1]$.  This stronger exchangeability assumption would hold, for example, under an unrealistically restrictive version of Figure \ref{dag} in which we removed all arrows from $Z_1,Z_2,Z_3$ into $S$.  Finally, note that our three identifying assumptions also give identification of the conditional counterfactual probability (\ref{truesub}) as a function of only the observed data.  We discuss this further in the next section where we consider estimating these functions and their time trends in practice where $Z_t$ is typically high-dimensional.

\subsection{Estimation}\label{estimation}
For the purposes of practical estimation and the assessment of trends in time $t$, it is useful to note that the expression (\ref{gform}) is algebraically equivalent to the weighted mean
\begin{equation}
\mbox{E}\left[\frac{Y_tS_t}{\Pr[S_t=1|Z_t]}\right]\label{weighted}
\end{equation}
where $\mbox{E}[]$ denotes expectation (population mean).  This weighted representation suggests a straightforward approach to estimating trends in the true underlying $\Pr[Y^{s_t=1}_t=1]$ via a weighted extension of regression modeling for estimating trends in the chance of a positive test $\Pr[Y_t=1|S_t=1]$ we considered in Section \ref{obsdata}.  In particular, we might use a weighted implementation of generalized estimating equation methodology \citep{repeatedweights} to estimate trends in (\ref{weighted}) as a function of $t$, which will give trends in $\Pr[Y^{s_t=1}_t=1]$ under our identifying conditions.  As above, we might make restrictive assumptions about these trends, such as 
\begin{equation*}
\Pr[Y^{s_t=1}_t=1]=\mbox{exp}\{\psi_0+\psi_1t\}
\end{equation*}
Analogously, provided this regression model is correctly specified, a value of $\psi_1>0$ would be consistent with an increasing trend in the true underlying disease prevalence under our three identifying assumptions.  By contrast, a value of $\psi=0$ would be consistent with no change in the underlying prevalence.   This model is simple because it assumes that $\frac{\Pr[Y^{s_t=1}_t=1]}{\Pr[Y^{s_{t-1}=1}_{t-1}=1]}$, is the same for all $t$.  Again, more complex models that allow more flexible functions of $t$ can also be considered (e.g. higher order polynomials, splines).  A completely flexible model would include separate indicator terms for each interval $t$ over the time period.  This would allow $\frac{\Pr[Y^{s_t=1}_t=1]}{\Pr[Y^{s_{t-1}=1}_{t-1}=1]}$  to differ for all $t$.  The weights in this implementation will take the value 0 for all individuals who did not receive a test.  For individuals who did receive a test, the weight is defined as the inverse probability of receiving a test in the interval $t$ population given that individual's level of the covariates $Z_t$. This probability might itself be estimated via regression modeling (e.g. a logistic regression model) when $Z_t$ is high-dimensional.   Note that all individuals, including those who did not receive a test, will contribute to the estimation of $\Pr[S_t=1|Z_t]$ for the weight denominator.  Stabilized weights may alternatively be used that additionally multiply the weight numerator by an estimate of $\Pr[S_t=1]$.  The use of stabilized weights may increase precision \citep{msmref}.  

If interest is in trends in the conditional prevalence $\Pr[Y^{s_t=1}_t=1|V_t]$ for some subset $V_t$ of the baseline measured selection factors $Z_t$, then either (i) the regression model fit can be restricted to one level of $V_t$ of interest (e.g. younger men) or (ii) the regression model can include terms for $V_t$ in addition to $t$. A simple, yet restrictive, example for $V_t$ selected to only include the covariate ``age'' taking the value 1 if an individual is below age 25 and 0 otherwise is 
\begin{equation*}
\Pr[Y^{s_t=1}_t=1|age]=\mbox{exp}\{\psi_0+\psi_1t+\psi_2age\}
\end{equation*}
This model is restrictive, not only in the assumption of linearity in $t$ but also in that it assumes that trends are the same regardless of the level of (discretized) age.  This assumption might be relaxed by including an interaction term between $t$ and the covariate ``age''.  When we consider trends in  $\Pr[Y^{s_t=1}_t=1|V_t]$ which conditions on $V_t$, stabilized weights may be used that multiply the weight numerator by an estimate of  $\Pr[S_t=1|V_t]$. Finally, if we selected $V_t$ to include all selection factors $Z_t$, then trends in $\Pr[Y^{s_t=1}_t=1|V_t]=\Pr[Y^{s_t=1}_t=1|Z_t]$ may be estimated via unweighted regression modeling but with additional inclusion of all covariates $Z_t$ in the model along with $t$.  

\section{More testing among the diseased?}
In the previous section we outlined untestable assumptions under which we may identify and estimate changes in underlying disease prevalence over time where we can define this prevalence by the counterfactual quantity $\Pr[Y^{s_t=1}=1]$.   Note that, by probability rules, we can write the overall chance of a positive test $\Pr[Y_t=1,S_t=1]$ as the product $\Pr[S_t=1|Y_t=1]\Pr[Y_t=1]$ where $\Pr[Y_t=1]=\Pr[Y^{s_t=1}=1]$, the underlying disease prevalence, under our assumption that screening in interval $t$ cannot affect disease status in that interval (for each individual, his outcome in the actual world in interval $t$ is equal to the outcome he would have had if, possibly contrary to fact, he was screened in that interval).  The other probability in this product $\Pr[S_t=1|Y_t=1]$ is the chance of receiving a test among those with disease.  An increase in this quantity over time means that diseased people have an increasing chance of their disease being identified; a likely result of increased and targeted screening efforts.  

Trends in $\Pr[S_t=1|Y_t=1]$ might be assessed by trends in $\frac{\Pr[Y_t=1,S_t=1]}{\Pr[Y_t=1]}$.  By above arguments, the numerator is directly observed and the denominator is identifiable under the outlined assumptions.  In the special case where marginal exchangeability holds then we have the equality $\Pr[S_t=1|Y_t=1]=\Pr[S_t=1]$ such that trends in this quantity can be directly estimated as in Section \ref{obsdata}.  However, as above, marginal exchangeability is not consistent with the causal DAG in Figure \ref{dag} (and will not generally be realistic) due to the unblocked backdoor paths through measured and unmeasured covariates connecting $Y_t $ and $S_t$.  However, under conditional exchangeability, we can conclude that $\Pr[S_t=1|Y_t=1,Z_t]=\Pr[S_t=1|Z_t]$ which can be directly estimated.

\section{Conclusions}\label{discussion}
In this report, we considered an approach to understanding assumptions under which we can evaluate using only observed variables whether increases in testing positivity over time are due only to increased testing among the disease, increased population disease prevalence or both.  We showed how causal diagrams can be useful for understanding these assumptions and for determining an analytic approach. 

\section{Acknowledgements}
The authors thank Adam M. Young for helpful discussions.  This activity was funded by the Division of STD Prevention, Centers for Disease Control and Prevention, through the STD Surveillance Network (CDC-RFA-PS13-1306). Support was also provided by the Massachusetts Department of Public Health.  JLM also received funding from National Institute of Allergy and Infectious Diseases (K01 AI122853).

\bibliographystyle{plainnat}
\bibliography{refs}

\end{document}